\documentclass[pra,a4paper,onecolumn,preprint,nofootinbib]{revtex4}
\usepackage{graphicx}
\usepackage{amsmath}

\def\inbar{\,\vrule height1.5ex width.4pt depth0pt}
\def\IR{\relax{\rm I\kern-.18em R}}
\def\IC{\relax\hbox{$\inbar\kern-.3em{\rm C}$}}

\newcommand{\eq}{\begin{equation}}
\newcommand{\fine}{\end{equation}}
\newcommand{\tint}{\textstyle\int}
\newcommand{\tiint}{\textstyle\iint}
\newcommand{\QATOP}[2]{\genfrac{}{}{0pt}{}{#1}{#2}}
\input{tcilatex}
\begin{document}

\title{Solving the Quantum Nonlocality Enigma by Weyl's Conformal
Geometrodynamics.}
\author{Enrico Santamato}
\email{enrico.santamato@na.infn.it}
\affiliation{Dipartimento di Fisica, Universit\`{a} Federico II, Napoli 80126}
\affiliation{CINSM - Consorzio Nazionale Interuniversitario per la Struttura della Materia}
\author{Francesco De Martini}
\affiliation{Accademia dei Lincei, via della Lungara 10, I-00165 Roma, Italy.\\
Dipartimento di Fisica, Universit\`{a} La Sapienza, I-00185 Roma, Italy.}
\email{francesco.demartini@uniroma1.it}
\date{\today}
\begin{abstract}
Since the 1935 proposal by Einstein Podolsky and Rosen the riddle of
nonlocality, today demonstrated by innumerable experiments, has been a cause
of concern and confusion within the debate over the foundations of quantum
mechanics. The present paper tackles the problem by a non relativistic
approach based on the Weyl's conformal differential geometry applied to the
Hamilton-Jacobi solution of the dynamical problem of two entangled spin 1/2
particles. It is found that the nonlocality rests on the entanglement of the
spin internal variables, playing the role of "hidden variables". At the
end, the violation of the Bell inequalities is demonstrated without recourse
to the common nonlocality paradigm. A discussion over the role of the "%
\textit{internal space"} of any entangled dynamical system involves deep
conceptual issues, such the \textit{indeterminism} of quantum mechanics and
explores the in principle limitations to any exact dynamical theory when
truly "hidden variables" are present. Because of the underlying geometrical
foundations linking necessarily gravitation and quantum mechanics, the theory
presented in this work may be considered to belong to the unifying
"quantum gravity" scenario.
\end{abstract}

\maketitle

\section{Introduction}

Since the 1935 publication of the famous paper by Einstein -- Podolsky --
Rosen (EPR), the awkward coexistence within the quantum lexicon of the
contradictory terms \textquotedblleft \textit{locality}\textquotedblright\
and \textquotedblleft \textit{nonlocality}\textquotedblright\ as primary
attributes to quantum mechanics (QM)\ has been a cause of concern and
confusion within the debate over the foundations of this central branch of
modern Science~\cite{EPR35}. In particular, even today this paradigmatic
conundrum keeps eliciting animated philosophical quarrels. For instance, an
extended literature consisting of articles and books is produced today by
eminent quantum field theorists endorsing the \textquotedblleft
local\textquotedblright\ side of the dilemma by a \textquotedblleft \textit{%
principle of locality}\textquotedblright\ based on the premise that quantum
observables measured in mutually spacelike separated regions commute with
one other~\cite{Haag,Dop1,Dop2}. On the other hand, the confirmation by
today innumerable experiments, following the first one by Alain Aspect and
coworkers, of the violation of the Bell inequalities emphasizes the dramatic
content of the dispute~\cite{Aspect,Bell,Redhead}. By referring to the
implications of Relativity with the nonlocal EPR correlations the
philosopher Tim Maudlin writes: \textquotedblleft \textit{One way or
another, God has played us a nasty trick. The voice of Nature has always
been faint, but in this case it speaks of riddles and mumbles as well..."}
~\cite{Maudlin}. Indeed, as it has been well known for three decades, the
experimental violation of the Bell inequalities implies the existence of
quite \textquotedblleft mysterious\textquotedblright\ nonlocal correlations
linking the outcomes of the measurements carried out over two spatially
distant particles. Moreover, in spite of these correlations any superluminal
transfer of useful information is found to be forbidden according to a
\textquotedblleft no-signalling theorem\textquotedblright . Recently this
was even tested experimentally~\cite{DeMa07}.\\

Aimed at a clarification of the problem, the present article
tackles the well known EPR scheme and explains by an exact analysis the
violation of the Bell's inequalities through a non relativistic approach,
for simplicity. Two equal\textit{\ }spin-1/2\textit{\ }particles $A$ and $B$%
, e.g. two neutrons, propagate in opposite directions along the spatial $y$%
-axis $(\overrightarrow{y})$ of the Laboratory with a velocity $v\ll c$
towards two spatially separate measurement devices, dubbed Alice and Bob,
who measure the spin of $A$ and $B$, respectively. Each apparatus, measuring
the particle $A$ (or $B$), consists of a standard Stern-Gerlach (SGA) device
followed by a couple of particle detectors (D) that, being rigidly connected to
SGA, can be oriented with it by a rotation in the $\overrightarrow{x}$-$%
\overrightarrow{z}\ $plane at the corresponding angles$\ \theta _{A}$ (or $%
\theta _{B}$) taken respect to $\overrightarrow{z}$. Accordingly, $%
\overrightarrow{\theta _{A}}$ and $\overrightarrow{\theta _{B}}$ denote the
orientation axes of SGA$_{A}$ and SGA$_{B}$~\cite{Redhead}.

\section{The non relativistic quantum top and Weyl's curvature}

Keeping the validity of the \textquotedblleft principle of
locality\textquotedblright\ which eventually refers to the final measurement
on the spins, it appears clear that the main problem implied by\ quantum
nonlocality resides with the very \textquotedblleft
ontological\textquotedblright\ nature of the wavefunction $\psi (A,B)\ $that
links the two particles since their common spatial origin. This problem
involves the "completeness" status of $\psi $, its\ deep implications with
the "Schr\"{o}dinger Cat" Paradox and possibly the solutions offered by
several exotic theories, e.g. the one based on the "$\psi $ spontaneous
collapse"~\cite{GRW}, the Everett's\ "Many Worlds"~\cite{Dewitt} or the
Albert-Loewer "Many Minds" interpretations~\cite{Albert}. Indeed, the quantum
correlation affecting any particle measurement, e.g. the outcome obtained
by Bob once the corresponding one has been obtained by Alice (or viceversa),
is a factual event, implying a temporary (or permanent) mechanical (or
electrical) change of the very structure of a physical measuring device. Of
course, a structural change cannot be achieved if the \textquotedblleft
link\textquotedblright\ provided by $\psi (A,B)\ $were a purely
informational entity, as assumed by a fairly large number of scientists. In
facts, any measurement outcome or any probability being a "number", cannot
by itself determine a structural change on any physical apparatus. As
already stressed in previous papers, this, or similar problems strongly
suggest that the wavefunction is not a mere mathematical entity but consists
of a physical \textquotedblleft field\textquotedblright\ and, more
precisely, as we shall see, a "gauge field" acting within a quantum theory
based on Weyl's conformal differential geometry~\cite{Weyl}. Accordingly, the
present theory of the EPR\ process is Weyl-conformally invariant~\cite%
{Santa84,demartini11}.

Let's consider, within a "classical" framework, a single particle, say $A$,
consisting of a spinning spherical top of mass $m$ and inertial moment $%
I_{c}=ma^{2}$, $a$ being the top's gyration radius\footnote{The radius $a$
is not determined in the present theory; however, when the theory is extended
to the relativistic framework, the radius $a$ is fixed by the theory itself
to be of the order of the particle Compton wavelength~\cite{demartini11}.}. The top
configuration space, with dimensions $n=6$, is given by the direct product of the
"\textit{external space}" of the Laboratory $\{x^{i}\}$, spanned by the
top's center of mass coordinates $x^{i}=$ $\left\{ x,y,z\right\} $, and of
the "\textit{internal space}" $\left\{ \zeta ^{\alpha }\right\} $ spanned by
the Euler's angles: $\zeta ^{\alpha }=\left\{ \alpha ,\beta ,\gamma \right\}
$ i.e. the top's \textquotedblleft \textit{internal variables}%
\textquotedblright\ fixing its orientation in space $\left\{ x^{i}\right\} $%
. The coordinates in the configuration space $\left\{ q^{\mu }\right\} $ are
then $q^{\mu }=$ $\left\{ x,y,z,\alpha ,\beta ,\gamma \right\} $, $(\mu
=1,...6)$. The kinetic energy of the top is: $K\ $=$\
{\frac12}%
(mv^{2}+I_{c}\omega ^{2})\ $= $%
{\frac12}%
\ mg_{\mu \nu }\dot{q}^{\mu }\dot{q}^{v}~$where the spatial components of
the velocity vectors are: $v^{i}=\left\{ \dot{x},\dot{y},\dot{z}\right\} $
and $\omega ^{i}$ $\doteq \ \lambda _{\alpha }^{i}\dot{\zeta}^{\alpha }\ $= $%
\left\{ -\dot{\beta}\sin \alpha +\dot{\gamma}\cos \alpha \sin \beta \text{, }%
\dot{\beta}\cos \alpha +\dot{\gamma}\sin \alpha \sin \beta \text{,}\ \dot{%
\alpha}+\dot{\gamma}\cos \beta \right\} $~\cite{Edmonds}. The tensor $g_{\mu
\nu }$,$\ $with $\det $er$\min $ant: $g=a^{6}\sin ^{2}\beta $, has a
diagonal 2-block form, i.e.: (a) a $3\times 3$ block: $g_{ij\ }=\delta _{ij}$
the diagonal Euclidean metric of the flat $\left\{ x^{i}\right\} $ space,
(b) a $3\times 3$ block $g_{\alpha \beta }=a^{2}\gamma _{\alpha \beta }$
with $\gamma _{\alpha \beta }$ the symmetric, nondiagonal Euler
metric tensor of the internal space $\left\{ \zeta ^{\alpha }\right\} $. The
internal metric $g_{\alpha \beta }$ exhibits a constant Riemann curvature: $%
R=3/(2a^{2})$. The quantities $\lambda _{\alpha }^{i}$ introduced in the
given expression of $\omega ^{i}$ can be considered as the parameters of a
set of three congruences in the internal space allowing the Euler metric $%
\gamma _{\alpha \beta }$ to be written in the dyadic form $\gamma _{\alpha
\beta }=\lambda _{\alpha }^{i}\lambda _{\beta }^{i}$ together with its
inverse $\gamma ^{\alpha \beta }=\mu _{i}^{\alpha }\mu _{i}^{\beta }$, where
$\mu _{i}^{\alpha }\lambda _{\beta }^{i}=\delta _{\beta }^{\alpha }$ and
$\mu _{i}^{\alpha }$ are the momenta of the congruences. The spin-1/2
operators components $\hat{s}_{i}=(\hbar /2)\hat{\sigma}_{i}$ on the spatial
axes $\left\{ x^{i}\right\} $ ($\hat{\sigma}_{i}$ are Pauli's operators) are
introduced as derivatives along line arcs: $\hat{s}_{i}\doteq -i\hbar \mu
_{i}^{\alpha }\partial _{\alpha }$. The standard spin commutation relations
hold: $\left[ \hat{\sigma}_{i},\hat{\sigma}_{j}\right] =2i\epsilon _{ijk}%
\hat{\sigma}_{k}$, as it may be checked by a direct calculation~\cite{Sakurai}.

The top configuration space is now endowed with the Weyl's connection
implied by the parallel transport of vectors of differential
geometry $\ \Gamma _{\mu \nu }^{\sigma }\ =-\left\{ \QATOP{\sigma }{\mu \nu }%
\right\} +\delta _{\mu }^{\sigma }\ \phi _{\nu }\ +\delta _{\nu }^{\sigma }\
\phi _{\mu }\ +g_{\mu \nu }\ \phi ^{\sigma }\ $where $g_{\mu \nu }\ $is the
metric tensor, $\left\{ \QATOP{\sigma }{\mu \nu }\right\} \ $\ the
Christoffel symbols\ and \ $\phi _{\mu }=\partial _{\mu }\phi $ \ is a
Weyl's vector assumed to be integrable, i.e. a gradient of a \textit{Weyl's
scalar potential}\ $\phi $~\cite{Weyl,Weinberg}. The quantity \ $\phi _{\mu }$
has been identified as a cosmological "\textit{World vector potential}" by
Peter G. Bergmann, the renowned Einstein's scholar and collaborator~\cite%
{Bergmann}. The dynamics of the top is realized considering the Lagrangian $L$
formed by adding to the kinetic energy $K$ a potential proportional to the Weyl
curvature: $L=K-\frac{\xi \hbar ^{2}}{m} R_{W}$ , where
$R_{W}=R+(n-1)[(2/\sqrt{g})\partial_{\mu }(\sqrt{g}g^{\mu \nu }\phi _{\nu })
-(n-2)g^{\mu \nu }\phi _{\mu }\phi _{\nu}]$. In the last equations, $R$ is the
Riemann curvature and, furthermore, $\xi $ is a numerical coupling parameter given by
$\xi=[(n-2)/8(n-1)]=$\ $1/10.$ The constant $\hbar^{2}$ implies that in the
present theory the potential $R_{W}$ is the source of all quantum features
of the system. The Hamilton-Jacobi equation (HJE) of the top's dynamics
is: $-\partial _{t}S=\frac{1}{2m}\ g^{\mu \nu }\partial _{\mu }S\
\partial _{v}S+\frac{\xi \hbar^{2}}{m}R_{W}$, where the "Action" $S$is the
Hamilton's principal function. The trajectories of the top in
the configuration space are given by the generalized velocities $v^{\mu }=%
\dot{q}^{\mu }=\frac{1}{m}g^{\mu \nu }p_{\nu }$, where $p_{\mu }=\partial
_{\mu }S$ are the generalized momenta. In order to determine the potential $%
\phi $, the HJE\ must be solved consistently with the \textit{continuity
equation} $\partial _{t}\rho +\frac{1}{\sqrt{g}}\partial _{\mu }(\sqrt{g}\
\rho v^{\mu })=0$, where the "density" is related to the Weyl potential $%
\phi $ by $\rho =A\exp [-(n-2)\phi ]$, $A$ being a normalization constant.
The last equation can be used to express the Weyl curvature $R_{W}$ \ in
terms of $\rho $ obtaining $R_{W}=R+[(n-1)/(n-2)][g^{\mu \nu
}\partial _{\mu }\rho \ \partial _{v}\rho /\rho ^{2}-(2/\rho \sqrt{g%
})\partial _{\mu }(\sqrt{g}g^{\mu \nu }\partial _{\nu }\rho )]$. Previous
articles reported a complete quantum solution of the present problem in a
fully relativistic framework, leading to a new, "ab initio" derivation of
the Dirac's equations~\cite{demartini11}. The key result of this approach is
that, unlike the metric which is fixed, Weyl's potential $\phi $, as well as
Weyl's curvature $R_{W}$, are determined by the top's motion. This motion,
in turn, is affected by $R_{W}$: a typical self-reacting geometrodynamical
process well known in the context of General Relativity~\cite{Wheeler}. In
other words, owing to the self-effect of $R_{W}\ $the single, apparently
isolated spinning particle can never be considered "free": as we shall see
soon, this unavoidable self-interaction is the basic geometrical origin of
the quantum nonlocality. For space limitations we report here the final
results of the solution adapted to the present nonrelativistic approach for
"free" particles~\cite{demartini11}. By means of the \textit{ansatz:} $\psi
(q^{\mu },t)=\sqrt{\rho (q^{\mu },t)}\exp [iS(q^{\mu },t)/\hbar ]$, here
interpreted as a "scalar wave function" satisfying Born's rule: $\rho
=\left\vert \psi \right\vert ^{2}$, the coupled problem implied by the
continuity and by the HJE equations is fully linearized - this is the
startling key result of the overall theory - and cast in the form of the
standard first-quantization theory based on the Schr\"{o}dinger-De-Rahm wave
equation:
\begin{equation}
i\hbar \partial _{t}\psi =-\frac{\hbar ^{2}}{2m\sqrt{g}}\partial _{\mu }(%
\sqrt{g}g^{\mu \nu }\partial _{\nu }\psi )+\frac{\xi \hbar ^{2}}{m}R\psi ,
\label{eq:wavequ}
\end{equation}%
where $R=3/(2a^{2})$ is the constant Riemann scalar curvature
calculated from the metric $g_{\mu \nu }$. Because of the linearity, this
wave equation is adopted, as usual, to describe the dynamical vector
evolution of the quantum state of the system in the standard Hilbert space%
~\cite{VonNeu}. The ensuing theory reproduces the standard quantum theory in
all formal details, in spite of \ the new dynamical interpretation of $\psi
(q^{\mu })$. As a quite remarkable feature, the Weyl's vector $\phi _{\mu
}(q^{\mu })$ and potential $\phi (q^{\mu })$ formally disappear from the
wave equation as they are kept hidden in the very definitions of $\rho
(q^{\mu })$ and $S(q^{\mu })$: this may explain why this or a similar theory
based on conformal Weyl's symmetry was never previously formulated. In facts,
the starting Lagrangian $L$ and all relevant equations occurring
in the theory are invariant under the Weyl's gauge transformations $g_{\mu
\nu }\rightarrow \lambda $ $g_{\mu \nu }$, $\phi _{\mu }\rightarrow \phi
_{\mu }+(1/2\lambda )\partial _{\mu }\lambda $ provided the values of the
"Weyl's weight", $w\ $of the action $S$ and of the mass $m$ are: $w(S)=0$
and $w(m)=-1$, respectively. As a consequence the velocity fields $v^{\mu }$,
the particle trajectories, the scalar density $m\rho \sqrt{g}$ and current $%
j^{\mu }=m\rho \sqrt{g}v^{\mu }$ have weight: $w=0$ i.e. they are all Weyl
gauge invariant. Also the eigenvalues $E$ of the Schr\"{o}dinger-De-Rahm
equations given by: $i\hbar \partial _{t}\psi =E\psi $ are gauge invariant.
The particle mass is not gauge invariant, but the mass ratio is gauge
invariant, which is all we need to have well defined mass values\footnote{The
mass enters as a parameter in the present non relativistic approach,
which is carried out in the gauge where $m=\mathrm{const.}$;
however, when the theory is extended to the relativistic framework, the
particle mass is no longer an external parameter, but it becomes a consequence
of the theory itself~\cite{demartini11}.}.

The general solution of the wave equation (\ref{eq:wavequ}) for the spin
$\frac12$
particle can be cast in the form
\begin{equation}
\psi (q^{\mu },t)=e^{-i\Omega t}[D_{\uparrow }(\alpha ,\beta ,\gamma )\psi
_{1}(x,y,z,t)+D_{\downarrow }(\alpha ,\beta ,\gamma )\psi _{2}(x,y,z,t)],
\label{eq:psisol}
\end{equation}%
where $\Omega =21\hbar /(40ma^{2})$ , $D_{\uparrow }(\alpha ,\beta ,\gamma
)=e^{\frac{1}{2}i(\alpha +\gamma )}\cos \frac{\beta }{2}=(\hat{D}%
^{-1}(\alpha ,\beta ,\gamma ))_{11}$, $D_{\downarrow }(\alpha ,\beta ,\gamma
)=e^{-\frac{1}{2}i(\alpha -\gamma )}\sin \frac{\beta }{2}=(\hat{D}%
^{-1}(\alpha ,\beta ,\gamma ))_{12}$ are the entries of the inverse of the
Wigner's SU(2) matrix $\hat{D}(\alpha ,\beta ,\gamma )$ representing the
3D-space rotation $\hat{R}_{3}(\alpha ,\beta ,\gamma )$, and $\psi
_{1}(x,y,z,t)$ and $\psi _{2}(x,y,z,t)$ are solutions of the time-dependent
Schr\"{o}dinger equation of the \textit{free} particle with mass $m$~\cite{Edmonds}.
We notice once again that the particle appears to be free in the wave equation,
because the geometric self-action of the particle on itself due to the Weyl
curvature is completely hidden in the structure of the wavefunction $\psi$.
Under space rotation $\{x^{i}\}\rightarrow \hat{R}_{3}\{x^{i}\}$, the
wavefunction $\psi (q^{\mu },t)$ changes as a scalar field, which implies
that the fields $\psi _{1}(x,y,z,t)$ and $\psi _{2}(x,y,z,t)$ change as the
two components of the Pauli spinor $\tilde{\psi}(x,y,z,t)=\binom{\psi
_{1}(x,y,z,t)}{\psi _{2}(x,y,z,t)}$. The spinor components $\psi
_{1}(x,y,z,t)$ and $\psi _{2}(x,y,z,t)$ correspond to the usual quantum
states of the particle with spin aligned in the positive and negative
direction of the Laboratory $z$-axis, respectively. In this way, spin 1/2
fields obeying the free particle time-dependent Schr\"{o}dinger equation
appear naturally in the theory, although the overall conformal invariance
symmetry is lost when a purely spinorial approach is used. In a more
familiar notation, Eq. (\ref{eq:psisol}) is written as $\psi (q^{\mu })$ $%
\equiv \left\langle x,y,z,\alpha ,\beta ,\gamma \mid \psi
_{spin}(t)\right\rangle $, where $\left\vert \psi _{spin}(t)\right\rangle
=\left\vert \psi _{1}(t),\uparrow _{z}\right\rangle +\left\vert \psi
_{2}(t),\downarrow _{z}\right\rangle $. The normalization is taken as $%
\langle \psi _{spin}(t)\left\vert \psi _{spin}(t)\right\rangle $= $\int
dxdydz(|\psi _{1}(x,y,z,t)|^{2}+|\psi _{2}(x,y,z,t)|^{2})=1$, where the
integration over the Euler angles was carried out with the usual measure: $%
d\mu (\alpha ,\beta ,\gamma )=\frac{1}{4\pi ^{2}}\sin \beta \,d\alpha d\beta
d\gamma $.

\section{Quantum entanglement and Weyl's curvature}

The extension to two or more spinning particles is straightforward. In the
case of two spins $A$ and $B$, the configuration space has $n=12$ dimensions
and is the direct product of the configurations spaces of the two particles,
spanned by the generalized coordinates $q^{\mu }=(x_{A},y_{A},z_{A},\alpha
_{A},\beta _{A},\gamma _{A};x_{B},y_{B},z_{B},\alpha _{B},\beta _{B},\gamma
_{B})$ $(\mu =1,\ldots 12)$. The kinetic part of the total lagrangian $%
L_{AB} $ is the sum of the kinetic parts of the Lagrangians of the two
particles and the metric $g_{\mu\nu}$ induced in the configuration space has
still a block diagonal form where the space and angle degrees of freedom are
separated. The Riemann curvature of the metric $g_{\mu\nu}$ is the sum of the
Riemann curvature of the two single particle subspaces: its value, $%
R=3/a^{2}\ $\ doesn't contribute to any inter-particle correlations (we
assume equal mass and equal inertia moments for the two particles). However,
the Weyl curvature $R_{W}(q^{\mu })$ for the two particles does not split,
in general, into separate contributions. Indeed, besides the geometrical
self-action of each particle, an interaction between the two particles is
established. This interaction disappears only in a very particular case,
i.e. in absence of quantum entanglement, as we shall see shortly. We
realize, therefore, that the very true origin of all effects related to
quantum entanglement, including the Bell's inequalities violation, resides
in the not eliminable interaction due to the presence of the Weyl curvature
in the Lagrangian. Formally, the key point consists of the logarithmic
dependence of the Weyl potential $\phi $ on the quantity $\rho =|\psi |^{2}$
according to: $\phi (q^{\mu })=$ $-(n-2)^{-1}\ln (\rho );$ ($n=12$). In the
absence of entanglement, we have $\psi =\psi _{A}\psi _{B}$ and $\rho =\rho
_{A}\rho _{B}$, so that $\phi =\phi _{A}+\phi _{B}$, where $\phi _{A}$ (or $%
\phi _{B}$) depends on the coordinates of particle $A$ (or $B$) only.
Consequently, the Weyl curvature splits into $R_{W}=R_{W}(A)+R_{W}(B)$, the
solution of the HJE splits into $S=S_{A}+S_{B}$, the velocity field splits
into $v^{\mu }=v_{A}^{\mu }+v_{B}^{\mu }$ and the overall continuity
equation splits into two separated continuity equations. On the other hand,
when the wavefunction $\psi $ cannot be separated, the same nonseparability
occurs for $R_{W}$, $S$, the velocity field, and the continuity equation. We
emphasize once again that the underlying interaction due to the Weyl curvature is
manifest in the HJE of the present theory, but it is completely hidden in
the quantum wave equation, which merely reduces to the sum of the separate
wave equations of the two particles. It follows that we cannot ascertain the
presence of entanglement just looking at the wave equation itself: we
must look instead at the form of its solutions. At the level of the equations of the
theory, entanglement is unveiled as a true physical phenomenon due to a non
trivial space  Weyl's curvature only within the context of the present approach.
The key point is that different solutions of the wave equations lead to
different Weyl's curvatures and, hence, to a different interaction among the
orientational degrees of freedom of the two particles. As we shall see by a
paradigmatic example in the next Section, the very origin of quantum nonlocality
relies on this unavoidable orientational geometric interaction.

\section{The EPR scalar wavefunction for two spin 1/2 particles}

Let us consider the EPR rotational invariant, "singlet" quantum state of
the particles $A$ and $B$ in motion along the $y$-axis:
\begin{equation}
\left\vert \psi _{AB}\right\rangle =\frac{1}{\sqrt{2}}[\left\vert \uparrow
_{z}^{A},\downarrow _{z}^{B}\right\rangle -\left\vert \downarrow
_{z}^{A},\uparrow _{z}^{B}\right\rangle .  \label{eq:psiABket}
\end{equation}%
The scalar wavefunction corresponding to this state is easily found to be
\begin{eqnarray}
\psi _{AB}(q^{\mu },t) &\equiv &\frac{1}{\sqrt{2}}e^{i(\frac{\gamma
_{A}+\gamma _{B}}{2}-2\Omega t)}\left[ e^{-i\frac{\Delta \alpha }{2}}\cos
\frac{\beta _{A}}{2}\sin \frac{\beta _{B}}{2}-e^{i\frac{\Delta \alpha }{2}%
}\cos \frac{\beta _{B}}{2}\sin \frac{\beta _{A}}{2}\right] \times
\label{eq:psiABwave} \\
&&\times \psi _{A}(x_{A},y_{A},z_{A},t)\psi _{B}(x_{B},y_{B},z_{B},t),
\end{eqnarray}%
where $\Delta \alpha =(\alpha _{B}-\alpha _{A})$. We don't consider Pauli's
exclusion principle here, because the \textit{external-space} wavefunctions $%
\psi _{A}(x^{i})\ $and$\ \psi _{B}(x^{i})$ are supposed to be associated to
two well separated wavepackets at positions $y_{A}$\ and $y_{B}$\ on the $y$%
-axis, respectively. The action and the density associated to $\psi
_{AB}(q^{\mu },t)$\ are
\begin{equation}
\begin{array}{c}
S(q^{\mu },t)=\hbar \lbrack -2\Omega +\frac{\gamma _{A}+\gamma _{B}}{2}%
+\arctan \left( \csc \frac{\beta _{A}-\beta _{B}}{2}\sin \frac{\beta
_{A}+\beta _{B}}{2}\tan \frac{\alpha _{B}-\alpha _{A}}{2}\right) + \\
+\arg (\psi _{A}(x_{A}^{i}))+\arg (\psi _{B}(x_{B}^{i}))]%
\end{array}
\label{eq:SAB}
\end{equation}%
and
\begin{equation}
\rho (q^{\mu },t)=\frac{1}{4}\left\vert \psi _{A}(x_{A}^{i})\right\vert
^{2}\left\vert \psi _{B}(x_{B}^{i})\right\vert ^{2}[1-\cos \beta _{A}\cos
\beta _{B}-\cos (\Delta \alpha )\sin \beta _{A}\sin \beta _{B}].
\label{eq:rhoAB}
\end{equation}

The differential equations of motion $\dot{q}^{\mu }=(1/m)g^{\mu \nu
}\partial _{\nu }S$ derived from (\ref{eq:SAB}) splits into three decoupled
sets: one involving the center of mass coordinates of particle $A$ only, one
involving the center of mass coordinates of particle $B$ only, and a third
set involving the Euler's angles of both $A$ and $B$. As said, the last set
of equations cannot be decoupled because of quantum entanglement. The
presence of entanglement is also unveiled by the expression of the Weyl
curvature $R_{W}$ derived from Eq. (\ref{eq:rhoAB}):
\begin{eqnarray}
R_{W}^{(AB)} &=&\frac{48}{5a^{2}}+\frac{22}{5a^{2}(1-\cos \beta _{A}\cos
\beta _{B}-\cos \Delta \alpha \sin \beta _{A}\sin \beta _{B})}+  \notag \\
&&+R_{W}^{(A)}(x_{A},y_{A},z_{A},t)+R_{W}^{(B)}(x_{B},y_{B},z_{B},t)
\end{eqnarray}%
where $R_{W}^{(A)}$ and $R_{W}^{(B)}$ are the Weyl spacetime curvatures
associated with the fields $\psi _{A}$ and $\psi _{B}$, respectively. We see
again that the total Weyl curvature is equally splitted into three terms:
the curvatures $R_{W}^{(A)}$ and $R_{W}^{(B)}$, depending on the space-time
"external" coordinates and yielding the self-action of each particle on
itself, and the coupling term which depends on the Euler angles only. Unlike
the spacetime terms, this last term cannot be splitted into the sum of two
independent potentials acting on each particle and it is the responsible of
all phenomena related to quantum entanglement. In this way, the very nature
of entanglement is explained as originating from the residual coupling of
the orientational degrees of freedom of the two spins due to the presence of
the Weyl's curvature $R_{W}^{(AB)}$. This one is the origin of a
inter-particle coupling consisting of a real orientational force that one
particle exerts on the other. As said, this force originates from $%
R_{W}^{(AB)}$, which in turn originates from Weyl's potential $\phi _{AB}$,
which ultimately arises from the system's wavefunction: $\psi _{AB}$. This
last one then looses its meaning of a purely mathematical entity in favor
of the more pregnant concept of a physical field. To summarize, in the
presence of entanglement, the \textit{internal} \textit{coordinates }
$\{ \zeta _{A/B}^{\alpha }\} $, viz. the Euler angles of the tops $A$ and $B$,
cannot be disentangled \textit{irrespective} of the mutual
spatial distance separating the two travelling particles. Even if these ones
are space-like separated by a large distance $d$, an inter-particle coupling
independent of $d$ arises that cannot be eliminated and is
responsible for the \textit{nonlocal EPR\ correlations}. Furthermore, we
conjecture from the present nonrelativistic standpoint that the space-time
superluminality of the nonlocal correlations comes from the
geometrical independency, i.e. disconnectedness, of the internal and
external manifolds:$\left\{ x^{i}\right\} $ and $\left\{ \zeta ^{\alpha
}\right\} $. This is the key result of the present Article. Note that the
dynamical Euler's angle coupling disappears in the \textit{absence of
entanglement}, i.e. in the case of a "\textit{product-state}". For instance%
\textit{, }for the state: $\left\vert \uparrow _{z}^{A},\downarrow
_{z}^{B}\right\rangle $ the term depending on the Euler angles in the Weyl's
curvature is: $R_{W}$ $=$ $-\frac{11}{5a^{2}}\left( \frac{1}{1+\cos \beta
_{1}}+\frac{1}{1-\cos \beta _{2}}\right) $, i.e. the contributions of $A$
and $B $ are mutually independent and separable.

In summary, when the entanglement is present, in order to save the
completeness of the theory the dynamical phase-space of any quantum system
must consist of the tensor product of the "\textit{external
space-time}" and of the "\textit{internal space}" viz.\ the one
spanned by the \textit{internal variables}, which may be interpreted as (non
measurable) dynamical "\textit{hidden variables}". This appears to be at
variance with the methods of standard quantum theory where the internal
variables are commonly integrated away, e.g. within the process of
definition of the "spin", which is itself a measurable quantity, for
instance by a SGA device. In this respect, we may conjecture that the (often
necessary) overlooking of the \textit{internal space} leads to several
consequences of deep conceptual and philosophical relevance. For instance, we
believe that some relevant manifestations of "\textit{quantum
indeterminism}" as well as the "\textit{quantum nonlocality}" precisely
arise from an over-simplified treatment of the dynamical problem, i.e. from
the neglect or the lack of knowledge of the internal variables of the
system. All this may draw us into even more profound speculations. Since in
our World - or in our Universe since the Big-Bang - all objects, bodies or
elementary particles are entangled because of the continuous, enormous
wealth of mutual interactions, even the "external" space-time geometry of
Special or General Relativity cannot be assumed, in principle, as a reliable
background of a complete dynamical theory. In the limit, no dynamics, no
mechanics, no physics would be possible but for systems brought, ironically,
into an "unphysical" isolation condition, e.g. in the case of few isolated
ions confined in an electromagnetic Paul trap~\cite{Paul}. \ Of course it has
been known for centuries that the solution of any physical problem always
implies the adoption of an idealized "filtering" of approximate dynamical
conditions:\ e.g. a successful study of the motion of the moon around the
earth must neglect the effect of the far galaxies. However, the paradigmatic
entanglement case accounted for in the present Article resists to any
approximation. Indeed, as we have seen, the mere neglect in the theory of
the manifold made of the "internal variables" - classified as "hidden"
because assumed to be inaccessible to measurement - leads irreducibly to an
unconceivable result, i.e. to the riddle of "quantum nonlocality" revealed
by the quite "mysterious" violation of the Bell's inequalities.

\section{The meaning of quantum measurement}

To better understand why the internal variables are not directly accessible
to experiments, we need a closer view of \ how experiments are carried out
in the quantum world. In essence, any experimental apparatus designed to
measure some physical property of a quantum particle is made of two parts: a
"filtering" device which addresses the particle to the appropriate detector
channel according the possible values of the quantity to be measured (e.g. a
spin component) and one (or more) detectors able to register the arrival of
the particle. To fix the ideas, we consider here the particular case of the
measure of a spin 1/2 particle by a Stern-Gerlach (SGA) apparatus. The spin
component along the SGA axis can have two values, so we need two detectors $%
D_{u}$ and $D_{d}$ coupled to the "up" and "down" output channels of the
orientable SGA. Each detector measures the flux $\Phi $ of particles
entering its acceptance area $A$. Let's assume single particle detection.
Then this flux is given by $\Phi =\tint_{A}j^{\mu }n_{\mu
}dA=\tint_{\Sigma }\rho \sqrt{g}g^{\mu \nu }\partial _{\nu }Sn_{\mu }d\Sigma
$ extended to the hypersurface $\Sigma $ in the particle configuration space
with normal unit vector $n_{\mu }=n_{\mu }=\{\mathbf{n},0,0,0\}$ where $%
\mathbf{n}$\ is the usual 3D-normal to the detector area $A$. Let us assume
that the scalar wavefunction of the particle at the detector location has
its spacetime and angular parts factorized, i.e. $\psi =\psi
_{1}(x,y,z,t)\psi _{2}(\alpha ,\beta ,\gamma )$. Then $\rho =\rho
_{1}(x,y,z,t)\rho _{2}(\alpha ,\beta ,\gamma )$, $S=S_{1}(x,y,z,t)+S_{2}(%
\alpha ,\beta ,\gamma )$ and $\Phi =\tint_{A}\mathbf{j}\cdot \mathbf{n}%
dA\tint \rho _{2}(\alpha ,\beta ,\gamma )d\mu (\alpha ,\beta ,\gamma )$,
where $\mathbf{j}=\rho _{1}(x,y,z,t)\nabla S_{1}$. The particle flux $\Phi $
is the only quantity directly accessible to the detector and depends only on
the spacetime part $\psi _{1}(x,y,z,t)$ of the wavefunction. The Euler's
angles are integrated away for the simple reason that the detector is
located in the physical space-time. It is worth noting that the current
density $j^{\mu }$ and, hence, the flux $\Phi $ is Weyl-gauge invariant as
it must be for any quantity having a measurable value.

Let us consider now the role played by the filtering apparatus. Unlike the
detector, whose role is just to count particles, the filtering stage of the
experimental setup must be tailored on the quantity to be measured. In the
case of the SGA the filtering device is the spatial orientation of the
inhomogeneous magnetic field crossed by the particle's beam. In an ideal
filtering apparatus no particle is lost, so its action on the particle's
wavefunction is unitary. The role of the filter is to correlate the
spacetime path of the particle with the quantity to be measured (the spin
component, in our case) so to extract from the incident beam all particles
with a given value of the quantity (spin "up", for example) by addressing
them to the appropriate detector. The filter acts on the particle motion in
space-time only. But, as said before, there is a feedback between the
particle motion and the geometric curvature of the space, so that the
insertion of the filter changes not only the particle path in spacetime, but
also the overall geometry of the particle configuration space, because it
modifies its Weyl's curvature $R_{W}$ through $\phi _{\mu }$, the, according
to Bergmann, environmental \textit{World vector potential}~\cite{Bergmann}.
This mechanism is at the core of General Relativity: the change in the
motion or the addition of a massive body changes the geometry of the whole surrounding
space. In our present approach, both particle motion and space geometry are
encoded in the scalar wavefunction, which indeed changes under the action of
the "unitary", i.e. lossless, transformation introduced by the SGA filter.
Solving the full dynamical and geometric problem inside the SGA is a
difficult problem, but the asymptotic behavior of the scalar wavefunction
far from the SGA may be easily found. In this "far-field scattering
approximation", a uniformly polarized particle beam is transformed by a SGA
rotated at angle $\theta $ with respect to the $\overrightarrow{z}$-axis as
follows,
\begin{eqnarray}
&&[aD_{\uparrow }(\alpha ,\beta ,\gamma )+bD_{\downarrow }(\alpha ,\beta
,\gamma )]\psi (x,y,z,t)\overset{SGA}{\longrightarrow }  \label{eq:SGA} \nonumber\\
&&\left( a\cos \frac{\theta }{2}+b\sin \frac{\theta }{2}\right) \left(
D_{\uparrow }(\alpha ,\beta ,\gamma )\cos \frac{\theta }{2}+D_{\downarrow
}(\alpha ,\beta ,\gamma )\sin \frac{\theta }{2}\right) \psi
(x_{u},y_{u},z_{u},t)+ \nonumber\\
&&+\left( a\sin \frac{\theta }{2}-b\cos \frac{\theta }{2}\right) \left(
D_{\uparrow }(\alpha ,\beta ,\gamma )\sin \frac{\theta }{2}-D_{\downarrow
}(\alpha ,\beta ,\gamma )\cos \frac{\theta }{2}\right) \psi
(x_{d},y_{d},z_{d},t)
\end{eqnarray}%
where $a,b$ are arbitrary complex constants with $|a|^2+|b|^2=1$, and labels
"u" and "d" refer to the positions of the detectors located to
the up and down exit channels of the $\theta$-oriented SGA. The
experimental apparatus is arranged so that the wave packets $\psi
(x_{u},y_{u},z_{u},t)$ and $\psi (x_{d},y_{d},z_{d},t)$ have negligible
superposition so that each detector sees a wavefunction with space and
angular parts factorized. Thus, for example, the particle flux detected in
the "up" channel of the SGA is given by $\Phi _{u}P_{u}(\theta )$, where $%
\Phi _{u}$ is the particle flux on the detector and $P_{u}(\theta
)=\left\vert a\cos \frac{\theta }{2}+b\sin \frac{\theta }{2}\right\vert ^{2}$
is usually interpreted as the probability that the particle in the input
wavepacket is found with its spin along the "up" direction of the SGA.

What the filter does is to correlate the particle space-time trajectory with
the quantity to be measured. In the standard quantum mechanical language, we
may say that the filter introduces a controlled entanglement among the
quantity to be measured and the particle spacetime path (in the SGA case,
the spacetime degrees of freedom become entangled with the orientational
ones). However, the filter is configured so that the wavepackets arriving on
each detector ($D_{u}$ and $D_{d}$, in our case) are not superimposed, and
the (approximate) wavefunction seen by each detector is of the product form
as considered above. The last requirement ensures that the detected particle
flux $\Phi $ provides a correct measure (in the quantum sense) of the
measured quantity\footnote{It is precisely the lack of this condition which
prevents to use the SGA to measure the spin of electrons. A way to overcome
this fundamental limitation was proposed very recently~\cite{karimi12}.}.

\section{The EPR state and Bell inequalities}

Let's now turn our attention to the joint spin measurements of the EPR
entangled particles $A$ and $B$ described by Eq. (\ref{eq:psiABwave}). After
leaving the source, particles $A$ and $B$ travel towards two Stern Gerlach
setups, SGA$_{A}$ and SGA$_{B}$, respectively, located at Alice's and Bob's
stations on two distant sites along $\overrightarrow{y}$. As said before,
each SGA acts \textit{locally}, by a \textit{unitary} transformation, on the
particle spatial, i.e. \textit{external}, degrees of freedom by correlating
its exit direction of motion with the direction of its spin respect to the
SGA\ axis, rotated around $\overrightarrow{y}\ $at angle $\theta $, taken
respect to $\overrightarrow{z}$. Since we are dealing with
$\frac12$%
-spins, there are only two exit directions, either "up" or "down" available
to each particle which will be then finally registered by a corresponding
detector. Let's refer to the Alice's and Bob's detectors as $D_{Au}$, $%
D_{Ad} $,$\ D_{Bu}$,$\ D_{Bd}$ and let $\theta _{A}$ and $\theta _{B}$.the
angles of SGA$_{A}$ and SGA$_{B}$, respectively.$\ $Labels "u" or "d" refer
to the particle's exit directions from each SGA's. As said above, the
presence of the two SGA changes not only the trajectories of the two
particles, but also the Weyl curvature of their configuration space.
These changes are both encoded in the change of the
wavefunction $\psi _{AB}$ in Eq. (\ref{eq:psiABwave}). Near the source that
wavefunction remains approximately unchanged, but far beyond the spatial
positions of the two SGA's the paths of the particles acquire different
direction according to their spin so that near the locations of the
detectors the input wavefunction is transformed according to
\begin{eqnarray}
&&\psi _{AB}\overset{SGAs}{\longrightarrow }A_{u,u}\psi _{A}(\mathbf{r}%
_{Au}t)\psi _{B}(\mathbf{r}_{Bu},t)+A_{u,d}\psi _{A}(\mathbf{r}_{Au},t)\psi
_{B}(\mathbf{r}_{Bd},t)+ \nonumber \\
&&+A_{d,u}\psi _{A}(\mathbf{r}_{Ad},t)\psi _{B}(\mathbf{r}%
_{Bu},t)+A_{d,d}\psi _{A}(\mathbf{r}_{Ad},t)\psi _{B}(\mathbf{r}_{Bd},t)
\end{eqnarray}%
where $\mathbf{r}_{Au}$, $\mathbf{r}_{Ad}$, $\mathbf{r}_{Bu}$, $\mathbf{r}%
_{Bd}$  are the positions of the detectors and $A_{u,u}$, $A_{u,d}$, $A_{d,u}$,
$A_{d,d}$ are coefficients depending on the two particle Euler's
angles and on the angles $\theta _{A}$ and $\theta _{B}$ of SGA$_{A}$ and SGA%
$_{B}$, respectively. The coefficients $A\ $can be easily calculated by
applying Eq. (\ref{eq:SGA}):
\begin{subequations}
\begin{eqnarray}
A_{u,u} &=&\chi (t)\left( D_{\uparrow }(\alpha _{1},\beta _{1},\gamma
_{1})\cos \frac{\theta _{A}}{2}+D_{\downarrow }(\alpha _{1},\beta
_{1},\gamma _{1})\sin \frac{\theta _{A}}{2}\right) \times  \label{eq:A} \nonumber \\
&&\times \left( D_{\uparrow }(\alpha _{2},\beta _{2},\gamma _{2})\cos \frac{%
\theta _{B}}{2}+D_{\downarrow }(\alpha _{2},\beta _{2},\gamma _{2})\sin
\frac{\theta _{B}}{2}\right) \sin \Delta \vartheta \\
A_{u,d} &=&\chi (t)\left( D_{\uparrow }(\alpha _{1},\beta _{1},\gamma
_{1})\cos \frac{\theta _{A}}{2}+D_{\downarrow }(\alpha _{1},\beta
_{1},\gamma _{1})\sin \frac{\theta _{A}}{2}\right) \times \nonumber \\
&&\times \left( -D_{\uparrow }(\alpha _{2},\beta _{2},\gamma _{2})\sin \frac{%
\theta _{B}}{2}+D_{\downarrow }(\alpha _{2},\beta _{2},\gamma _{2})\cos
\frac{\theta _{B}}{2}\right) \cos \Delta \vartheta \\
A_{d,u} &=&\chi (t)\left( -D_{\uparrow }(\alpha _{1},\beta _{1},\gamma
_{1})\sin \frac{\theta _{A}}{2}+D_{\downarrow }(\alpha _{1},\beta
_{1},\gamma _{1})\cos \frac{\theta _{A}}{2}\right) \times \nonumber \\
&&\times \left( D_{\uparrow }(\alpha _{2},\beta _{2},\gamma _{2})\cos \frac{%
\theta _{B}}{2}+D_{\downarrow }(\alpha _{2},\beta _{2},\gamma _{2})\sin
\frac{\theta _{B}}{2}\right) \cos \Delta \vartheta \\
A_{d,d} &=&\chi (t)\left( -D_{\uparrow }(\alpha _{1},\beta _{1},\gamma
_{1})\sin \frac{\theta _{A}}{2}+D_{\downarrow }(\alpha _{1},\beta
_{1},\gamma _{1})\cos \frac{\theta _{A}}{2}\right) \times \nonumber \\
&&\times \left( -D_{\uparrow }(\alpha _{2},\beta _{2},\gamma _{2})\sin \frac{%
\theta _{B}}{2}+D_{\downarrow }(\alpha _{2},\beta _{2},\gamma _{2})\cos
\frac{\theta _{B}}{2}\right) \sin \Delta \vartheta
\end{eqnarray}
\end{subequations}
where: $\chi (t)\equiv \frac{1}{\sqrt{2}}e^{-2i\Omega t}$ and:$\ (\Delta
\vartheta )\equiv (\theta _{B}-\theta _{A})/2$.\ \ The coincidence rate are
given by the joint particle fluxes intercepted by the detectors, viz. $\Phi
_{i,j}(\theta _{A},\theta _{B})=\tiint \left\vert A_{ij}(\alpha _{1},\beta
_{1},\gamma _{1},\alpha _{2},\beta _{2},\gamma _{2};\theta _{A},\theta
_{B})\right\vert ^{2}d\mu (\alpha _{1},\beta _{1},\gamma _{1})d\mu (\alpha
_{2},\beta _{2},\gamma _{2})\tint \mathbf{j}_{i}\cdot \mathbf{n}%
_{i}dA_{i}\tint \mathbf{j}_{j}\cdot \mathbf{n}_{j}dA_{j}$, where $i,j=u,d$
and: $\mathbf{j}_{i}=\left\vert \psi _{A}(\mathbf{r}_{i},t)\right\vert
^{2}\nabla S_{A}(\mathbf{r}_{i},t)$, $\mathbf{j}_{j}=\left\vert \psi _{B}(%
\mathbf{r}_{j},t)\right\vert ^{2}\nabla S_{B}(\mathbf{r}_{j},t)$ are the
particle current densities at the detectors. A simple calculation shows
that if all particles falling into the detectors are counted, the
coincidence fluxes are given by $\Phi _{u,u}(\theta _{A},\theta _{B})=\Phi
_{d,d}(\theta _{A},\theta _{B})=\frac{1}{2}\sin ^{2}\left( \Delta \vartheta
\right) $ and $\Phi _{u,d}(\theta _{A},\theta _{B})=\Phi _{d,u}(\theta
_{A},\theta _{B})=\frac{1}{2}\cos ^{2}\left( \Delta \vartheta \right) $. The
coincidence fluxes $\Phi _{ij}$ are Weyl-invariant and can be experimentally
measured. Moreover, they are equal to the joint probabilities $%
P_{i,j}(\theta _{A},\theta _{B})$ associated with the EPR state (\ref%
{eq:psiABwave}), in full agreement with the standard quantum theory and
lead straightforwardly to the violation of the Bell's inequalities within
all appropriate experiments consisting of statistical measurements over
several choices of \ the angular quantity $(\Delta \vartheta )$, as shown by
many modern texts~\cite{Bell,Scully,Redhead,Maudlin}. For instance, Redhead
considers the inequality: $\ F(\Delta \vartheta )\equiv \left\vert 1+2\cos
(2\Delta \vartheta )-\cos (4\Delta \vartheta )\right\vert \leq 2$ \ which is
violated for all values of $(\Delta \vartheta )$ between $0%
{{}^\circ}%
$ and $45%
{{}^\circ}%
$. In summary, in order to attain the correct result the present theory
promoted the "hidden variables" to the status of "internal variables" of the
particle's relevant property: the "spin". We believe that these variables,
i.e. the Euler's angles, should be considered as a necessary dynamical
aspects of that fundamental quantum property. Note that the standard "hidden
variables" no-go theorem is not violated by our theory, because in
standard quantum theory the not trivial curved configuration space and the
feedback between space curvature and particle motion are absent~\cite%
{VonNeu}.

\section{Conclusions}

We have demonstrated that the quantum nonlocality
enigma, epitomized by the violation of the Bell's inequalities, may be
understood on the basis of a Weyl's conformal geometrodynamics. This result
was reached through a theory that bears several appealing properties and may
lead to far reaching consequences in modern physics. We summarize them as
follows:

1) The linear structure of the standard first quantization theory is fully
preserved, in any formal detail.

2) The quantum wavefunction acquires the precise meaning of a physical
quantum "\textit{Weyl's gauge field}" acting in a curved configurational
space.

3) A proper theoretical analysis of any quantum \textit{entanglement}
condition must involve the entire configurational space of the system
including the usual space-time of General Relativity as well as the
"internal coordinates" of the system. If entanglement is present and if
the internal coordinates are really "hidden", i.e. if they are absent in the
theory - as they are in standard quantum theory - severe limitations
may arise on the actual interpretation of any dynamical problem. There
physics may even be an impossible task, in principle, and paradoxes may
spring out. \ Indeed, in addition to "\textit{quantum nonlocality}", many
counterintuitive concepts of quantum mechanics, such as those related to
several aspects of "\textit{quantum indeterminism}" and of "\textit{quantum
counterfactuality}" may precisely arise from these theoretical limitations.
Which are indeed limitations to the human knowledge and understanding.

4) The "sinister", "disconcerting" and "discomforting" aspects of
entanglement were expressed right after the publication of the EPR paper
by a surprised and highly concerned Erwin Schr\"{o}dinger. Who also added:
"I would not call that one but rather \textit{the} characteristic trait
of quantum mechanics, the one that enforces the departure from the classical
lines of thought"~\cite{Sch35}.

5) By solving an utterly important enigma the present paper clarifies
\textit{the} - according to Schr\"{o}dinger - characteristic trait of quantum
mechanics. The adopted  theory is based on a necessary significant aspect
of the interplay between geometry and matter motion on which also rests the
modern theory of gravitation, i.e. General Relativity. Consequently, our
work may be considered to belong to a unifying theoretical scenario
linking necessarily gravitation and quantum mechanics. This is indeed the
long sought, paradigmatic "quantum gravity" scenario.
%

\end{document}